\def\IGR{IGR~J17497$-$2821}
\def\SWIFT{Swift~J1753.5$-$0127}
\def\ergcms{erg cm$^{-2}$ s$^{-1}$ }
\def\integral{{\it{INTEGRAL}}}
\def\rxte{{\it{RXTE}}}

\documentclass[]{aastex}
\usepackage{emulateapj5}
\usepackage{color}
\definecolor{red}{rgb}{0.7,0,0}
\definecolor{black}{rgb}{0,0,0}
\def\correc#1{{\textcolor{black}{#1}}}

\submitted{Accepted in ApJ Letters}
\shorttitle{\IGR\ observed by \rxte\ and ATCA}
\shortauthors{Rodriguez et al.}
\begin{document}
\title{The discovery outburst of the X-ray transient \IGR\ observed with \rxte\ and ATCA}

\author{J\'er\^ome Rodriguez\altaffilmark{1}, 
Marion Cadolle Bel\altaffilmark{2,1}, John A. Tomsick\altaffilmark{3,4}, 
St\'ephane Corbel\altaffilmark{1}, Catherine Brocksopp\altaffilmark{5}, 
Ada Paizis\altaffilmark{6}, Simon E. Shaw\altaffilmark{7,8}, 
 Arash Bodaghee\altaffilmark{8,9}}

\altaffiltext{1}{AIM Astrophysique Interaction Multi-\'echelles  
CEA Saclay, DSM/DAPNIA/SAp, F-91191 Gif sur Yvette, Cedex France}
\altaffiltext{2}{European Space Astronomy Centre, Villafranca del Castillo, 28629
Madrid Spain}
\altaffiltext{3}{Space Sciences Laboratory, 7 Gauss Way,
University of California, Berkeley, CA 94720,  USA}
\altaffiltext{4}{Center for Astrophysics and Space Sciences,
9500 Gilman Drive, Code 0424, University of California at San Diego, La Jolla, CA 92093, USA}
\altaffiltext{5}{Mullard Space Science Laboratory, University College London, Holmbury St. Mary, Dorking, Surrey RH5 6NT, UK}
\altaffiltext{6}{IASF Milano--INAF, Via Bassini 15, 20133 Milano, Italy}
\altaffiltext{7}{School of Physics and Astronomy, University of Southampton, SO17 1BJ, UK}
\altaffiltext{8}{INTEGRAL Science Data Centre, Chemin d'Ecogia, 16, 1290 Versoix, Switzerland}
\altaffiltext{9}{Observatoire Astronomique de l'Universit\'e de Gen\`eve, Chemin des Maillettes 
51, CH-1290 Sauverny, Switzerland}

\begin{abstract}
We report the results of a series of \rxte\ and ATCA observations of the 
recently-discovered X-ray transient  \IGR. Our 3-200 keV PCA+HEXTE spectral analysis shows
very little variations over a period of $\sim$10 days around the maximum of the outburst. 
\IGR\ is found in a typical Low Hard State (LHS) of X-ray binaries (XRB), well represented 
by an absorbed Comptonized spectrum with an iron edge at about 7 keV.
The high value of the absorption ($\sim4\times 10^{22}$ cm$^{-2}$) suggests 
that the source is located at a large  distance, either close to the Galactic center 
or beyond. The timing analysis shows no particular features, while the shape of the 
power density spectra is also typical of LHS of XRBs, with $\sim$36\% RMS variability. 
No radio counterpart is found down to a limit of  0.21 mJy at 4.80 GHz  and 8.64 GHz. 
\correc{Although the position of \IGR\ in the radio to X-ray flux diagram is well 
below the correlation usually observed in the LHS of black holes,} the comparison 
of its X-ray properties with those of other sources leads us to suggest that it 
is  a black hole candidate.
\end{abstract}

\section{Introduction}
Transient X-ray Binaries (XRB) are known to show a series of different X-ray 
spectral states during their outbursts. A certain number of these 
objects have, however, undergone outbursts during which they remained in the Low Hard 
State (LHS) \cite[e.g.][]{hynes00,brocksopp04,rodriguez06b, cadolle06a}. A strong correlation
exists between the radio and X-ray fluxes during this state \citep{corbel00,gallo03}.
This may indicate some influence of the radio jet in 
the X-ray domain. The study of the LHS should, therefore, reveal important clues 
to the connection between the accretion and ejection processes. Interestingly, 
black hole (BH) transients are more radio loud than neutron stars (NS) by a factor of 
$\sim$30 \citep{miggliari06} \correc{in the LHS}.\\
\indent \IGR\ was discovered with ISGRI on-board \integral\ on Sept. 17, 2006
 as a new hard XRB \citep{soldi06} \correc{at Galactic coordinates l=0.97$^\circ$, 
b=-0.46$^\circ$.} \correc{Given that the line of sight to the source passes close to 
the Galactic Center, and assuming a  distance of 8 kpc, \citet{kuulkers06} 
estimated a 2-100 keV (unabsorbed) luminosity of $\sim10^{37}$ erg/s. The position and  
luminosity strongly indicate \IGR\ is a Galactic XRB.
The preliminary spectral analysis of their \integral\ data led \citet{kuulkers06} to further 
suggest the source was in a LHS typical of BH and NS XRBs}. A follow up 
observation with  the Chandra X-ray Observatory allowed a fine X-ray position to 
be given at \correc{$\alpha=17^h 49^m 38.037^s$, $\delta=-28^\circ 21' 17.37''$ 
($\pm0.6''$, at 90\% confidence, Paizis et al. 2006). This refined X-ray position 
permitted \citet{paizis06}
to identify the probable optical/IR counterpart, and classify it as a red giant K-type
star making \IGR\ a Low Mass X-ray Binary (LMXB).}\\
\indent Soon after the discovery, we triggered our \rxte\ program P92016 as well as radio 
observations with the Australian Telescope Compact Array 
(ATCA), aiming to identify the nature of the system. 
We report here the results of those campaigns.

\section{Observations and Data Analysis}
\subsection{X-ray observations}
Our\rxte\ program has been divided into seven pointings, the details 
of which are reported in Table \ref{tab:log}. Fig. \ref{fig:Lite} shows the 
3--30 keV \rxte/PCA and 15--50 keV Swift/BAT light curves of \IGR\ over the outburst.
The PCA data were reduced in the same  way as in \citet{rodriguez03}, 
although with the latest version of the {\tt{HEASOFT}} software package. 
Since the source is located in a crowded 
area, standard background subtraction is not sufficient to remove the contribution
of other sources and the Galactic Bulge (about 10 mCrab between 2-10 keV, 
\citet{markwardt06}). In order to better estimate the sky background at the source 
position, we analyzed an \rxte\ observation of the nearby
BH Candidate XTE J1748-288 made at the end of its 1998 outburst, when the 
source flux was comparable with that of the Galactic background (Obs. Id 30185-01-20-00).
The best fitted model of this observation was then used as an extra background 
component for the spectral fittings of all our observations.\\
\indent For HEXTE, we extracted spectral data from the Cluster B 
unit only, since Cluster A  no longer obtains a background measurement because the rocking
has \correc{ceased}. We checked that ``plus'' and ``minus''
offset pointings gave similar spectra (indicating that no active sources 
contributed to the background) and combined the two pointings to obtain background 
spectra. The remaining processes 
for HEXTE reduction are similar to the procedures presented in \citet{rodriguez03}.
The PCA and HEXTE spectra of each observation were fitted simultaneously in 
{\tt{XSPEC}} v11.3.2t between 3 and 200 keV. A normalization constant 
was introduced to account for calibration uncertainties between 
PCA and the single HEXTE cluster.\\
\indent  High resolution light curves were extracted from Good Xenon and Event 
data with a time resolution of $\sim122\mu$s allowing us to study the temporal properties 
of the source up to 4096 Hz. In order to restrict the background effects at low and high 
energies, we restricted the extraction to the $\sim5$--$40$ keV range. Since QPOs and 
coherent pulsations are usually stronger at these energies 
\citep[e.g.][]{morgan97,rodriguez04,rodriguez06a}, this ensures a higher sensitivity to any
feature that might exist.  
We produced power density spectra (PDS) from each of the individual 
light curves between $\sim0.02$ and 4096 Hz. 
\subsection{Radio observations}
\IGR\ was observed on  Sept. 26 and 27, 2006, with the ATCA, 
located at Narrabri, Australia. The array was in the H75 configuration 
(allowing a baseline as long as 6 km)  with antennas 1 and 5 off-line 
due to maintenance. Observations 
were carried out simultaneously at 4.80 GHz (6 cm) and 8.64 GHz (3.5 cm), 
with a continuum bandwidth of 128 MHz, for a total of about 12.3 hours on source. 
The amplitude and band-pass calibrator was PKS~1934$-$638, and the antennas' 
gain and phase calibration, as well as the polarization leakage, were 
derived from regular observations of the
nearby calibrator TXS 1748$-$253. The editing, calibration, Fourier transformation, 
deconvolution, and image analysis 
were performed using the MIRIAD software package (Sault \& Killen 1998). 
No radio emission from \IGR\ was detected 
with ATCA on either date. Combining the two sets of observations 
resulted in the three sigma upper limits for \IGR\ of 0.21 mJy at 
both frequencies.

\section{Results of the X-ray analysis}
\begin{figure}
\epsscale{0.4}
\plotone{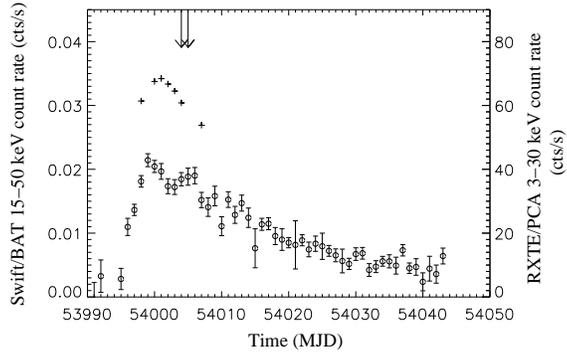}
\caption{15--50 keV Swift/BAT (open circles) and 3--30 keV \rxte/PCA (crosses) light curves
 of \IGR\ during the outburst studied in this paper. The vertical arrows represent 
the dates of our ATCA observations. For BAT 1 Crab $\sim0.23$ cts/s and $\sim1840$ cts/s for PCA.}
\label{fig:Lite}
\end{figure}

\begin{figure}
\epsscale{0.4}
\plotone{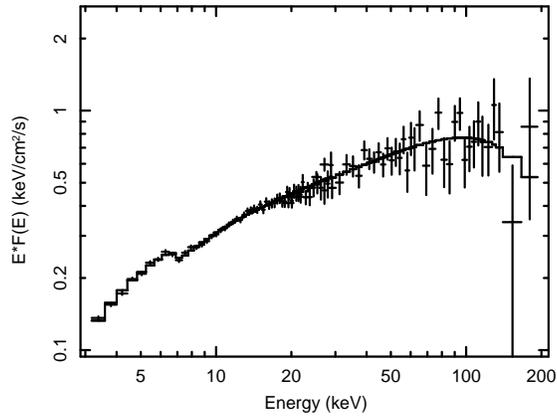}
\caption{3-200 keV \rxte\ spectrum of Obs. 7. The best model is superimposed 
as a line.}
\label{fig:spectrum}
\end{figure}

\begin{figure}
\epsscale{0.4}
\plotone{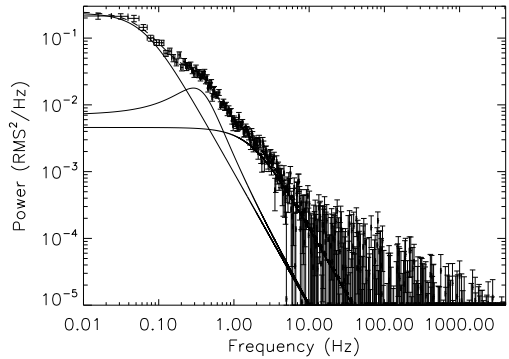}
\caption{White noise corrected PDS from Obs. 1. The continuous lines represents the 3 broad Lorentzians 
used to model the PDS. The dashed lines is the best model.}
\label{fig:pds}
\end{figure}

Fig. \ref{fig:Lite} shows the light curves seen by Swift/BAT and \rxte/PCA. 
 In about 5 days the 15-50 keV BAT flux increased by a factor greater than 5, and  
remained around 0.02 cts/s until MJD 54006. The BAT flux is not purely constant though, 
since it showed  the presence of two peaks.  After MJD 54006 
a significant decrease is seen (Fig. \ref{fig:Lite}). From the quality of the data, 
it is difficult to say whether the shape of the BAT light curve 
is FRED-like,  or if the decay is purely linear. In fact,  
fitting the decay with an exponential leads to an e-folding time of 16.9 days. \\
\indent The spectra were first fitted with a simple absorbed 
power law model. In all cases, the power law photon index was hard ($\sim1.56$).
Although the reduced $\chi^2$ was relatively acceptable (1.5 for 96 
degrees of freedom (DOF) for Obs. 1), a slight deviation 
was visible at high energy. Adding a cutoff improved the fit. An F-test
yielded a probability of 2.8$\times$10$^{-3}$ that the improvement 
was purely due to chance. In all observations except 4, 5 and 6 the cutoff 
was required at a high level of significance.  This may indicate that 
it was either truly absent in obs. 4, 5 and 6, or that the cutoff energy 
had increased closer to the upper boundary of our spectral analysis. Note that these 
three observations occur between the two peaks of the BAT light curve (Fig. \ref{fig:Lite}).\\
\indent Since a hard power law with an exponential cutoff is usually interpreted 
in XRBs as being due to thermal Comptonization, the pure phenomenological model was 
replaced by the {\tt{comptt}} model \citep{titarchuk94}. The same model was applied to all
observations to ease the comparison. 
In all cases, a hint for an iron edge at $\sim7$ keV was visible. Adding such a feature greatly improved 
the fits (1.7$\times$10$^{-9}$ chance improvement). The mention by \citet{itoh06} of such a 
feature in the {\it{Suzaku}} spectrum of the source lent further credibility to this detection.  
The optical depth of the Edge ranges from 0.08 to 0.12, while the 
reduced $\chi^2$ ranges from 0.97 to 1.14 for 93 DOF. All other  
fit parameters are reported in Table \ref{tab:fits}, while Fig. \ref{fig:spectrum} shows 
the   spectrum from Obs. 7.  The source spectral parameters were 
relatively stable
over the period of observations. Slight differences are found only for Obs. 1 \& 7, for which 
the electron temperatures seem higher, although poorly constrained.\\
\indent The 1-s bin PCA light curves extracted from the 7 observations showed no particular differences. 
 No obvious dips, eclipses, or X-ray bursts were 
visible\footnote{In the PCU 3 light curve of Obs. 2, however a large flare was 
detected. This was not confirmed while inspecting all other active PCUs. This flare 
seems to be purely instrumental.}. 
To quantify the degree of variability, we inspected each PDS 
individually. The variability of \IGR\ is dominated by frequencies lower than 1 Hz 
(Fig. \ref{fig:pds} for Obs. 1). We modeled the PDSs with the help 
of broad Lorentzians \citep{belloni02}. As in typical PDSs of XRBs, three such features
 are needed \citep{vanderklis06}, two with their 
centroid frozen to 0 and one representing the so called 
low frequency hump (LFH). Broad Lorentzians mimic a flat top component 
and power law decay above a break frequency. The analogue of the break frequency is 
given by $\Delta=\sqrt{(\sigma/2)^2+\nu_{0}^2}$, with $\sigma$ the full width at half
 maximum, and 
$\nu_0$ the centroid frequency \citep{belloni02}. The PDSs from the different observations
showed very little variations. The 3 components  were compatible (within the errors) 
with being constant through the outburst. As a reference the parameters we obtained  
for Obs. 1 
are $\Delta=0.06\pm0.01$ Hz, RMS$\sim13\pm2\%$, for the main component $\Delta=1.8\pm0.7$ 
Hz, RMS$=10\pm2\%$
for the second, and $\nu_0=0.30\pm0.07$ Hz, $\Delta=0.23\pm0.07$ Hz, RMS$=13\pm2\%$ for 
the LFH. 
The  \correc{total} RMS variability is then $\sim$36$\%$ RMS. The  $3\sigma$ upper 
limit for a 2 Hz FWHM QPO is 2\%. \\
\indent \correc{We searched for X-ray pulsations between $\sim0.2$ ms and 256 s in the 
unbinned PDSs. We do not detect any coherent pulsations in any of them. While 
below 1 Hz the significant level of aperiodic noise  renders our search less 
sensitive to the detection of a given feature, above 1 Hz a periodic signal would be more 
readily detectable out of the white noise. 
From the PDS of  Obs. 1, and with a 
frequency resolution of 3.9$\times10^{-3}$ Hz, we 
could calculate $3\sigma$ upper limits of $0.9\%$ between 1 and 4096 Hz. Below 1 Hz, 
using a similar approach (assuming here that a pulsation should appear out of 
the level of aperiodic noise, and using the ``flat top'' level below 0.3 Hz), 
we could calculate that the limit for a coherent pulsation 
ranges from $\sim2.4 \%$ at $0.4$ mHz to $\sim1\%$ at 1 Hz.}

\section{Discussion}
\begin{figure}
\epsscale{0.5}
\plotone{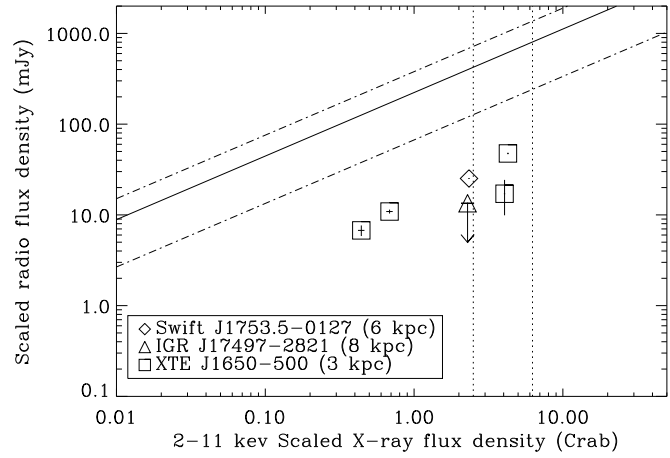}
\caption{Plot of the radio vs. X-ray fluxes for the BH XTE J1650$-$500 \citep{corbel04}, the BHC 
Swift J1753.5$-$0127 \citep{cadolle06a}, and \IGR\, with their (assumed) distances. All fluxes 
are normalized to a distance of 1 kpc. The range of correlated radio/X-ray fluxes is indicated 
by the dashed dot lines, while the continuous line indicates the best position. The vertical dotted 
lines indicate 2\% and 5\% of the Eddington luminosity.}
\label{fig:correl}
\end{figure}

We analyzed a set of \rxte\ observations of the newly-discovered source \IGR.
The PDSs of \IGR\ are typical of XRBs in the LHS \cite[e.g.][]{vanderklis06}. Although 
the general
shape of the PDSs makes it difficult to discriminate a NS from a BH \citep{vanderklis06}, 
the typical frequencies are believed to be lower in the case of BH (the ms
accreting pulsar IGR J00291+5934 being a recent counter example). In fact \citet{sunyaev00} 
observed that in the LHS for BH systems no significant signal is 
detected above $\sim50$ Hz, contrary to weakly magnetized NS. In \IGR\  
the lack of significant signal above $\sim$10 Hz, \correc{above which the level of variability 
is compatible with being purely due to Poisson noise}, and the high level of 
the flat top noise together with the low frequency of the break 
\correc{in the first Lorentzian 
component (0.06 Hz)} are reminiscent of BH systems. 
Generally speaking, and although this is not a proof, none of the characteristics that 
would make \IGR\ a definite NS (kHz QPOs, X-ray bursts, coherent pulsations) are observed. \\
\indent The source spectrum is well represented by a power law (with $\Gamma\sim 1.57$) 
convolved by interstellar absorption and a high-energy cutoff \correc{starting at  
$E_{\mathrm{cut}}\sim 50$ keV with a folding energy $E_{\mathrm{fold}}\sim190$ keV}, which are  
all typical of XRBs in the LHS.  
This is usually interpreted as due to thermal Comptonization of soft X-ray photons 
on a population of energetic electrons. Whether these electrons form a ``corona'' or the base of 
a compact jet is subject to debate \correc{\citep[e.g.][]{markoff05}}. 
A thermal Comptonization model represents the spectra well.  
The high value of the electron temperature and the optical depth, suggest 
that the compact object  probably  a BH. Indeed, with $E_{cut}=3\times kT_{e}$, we obtain 
an  equivalent cutoff energy of about 100-120 keV which is similar to e.g. Cyg X-1 while in the
LHS \citep{cadolle06}, or XTE J1550$-$564 during its 2002 outburst \citep{belloni02b}), two well
known BHs. In addition, even if some extremely hard NS Atoll sources can show similar spectra,  
the values 
of $\tau$ obtained in those cases are systematically higher \citep{barret01} than those of \IGR.\\
\indent The (unabsorbed) 1-20 and 20-200 keV luminosities are 
$1.1\times10^{37}\times (d/8kpc)^2$erg/s and 
$2.1\times10^{37}\times (d/8kpc)^2$erg/s. At the distance of the Galactic Center,  these 
luminosities place the source outside the so-called burster box \citep{barret00}, 
in fact very close to Cyg X-1 in its LHS. 
Although several NS sources have been seen to lie outside this box, they all have their
 20--200 keV 
luminosity lower than $\sim10^{37}\times$erg/s. 
In this respect, if \IGR\ is located at a large distance, as possibly 
suggested by the high value of the equivalent absorption column density, its hard luminosity
strongly suggests that it contains a BH. \\
\indent  Fig. \ref{fig:correl} shows
the position of \IGR\ w.r.t. the radio-X-ray correlation. \IGR\ lies 
below the expected correlation for BHs.
Fig \ref{fig:correl} \citep[see][]{corbel04,cadolle06a} shows that some BH(C)s have been 
observed to lie below the expected correlation. The 
non-detection of \IGR\ in radio is therefore not an argument against a BH primary. 
Given the results of our X-ray spectral analysis, we conclude that \IGR\ is a BHC in its
LHS. \correc{The observations of sources for which the correlation between the radio 
and X-ray fluxes 
is not observed suggest that, contrary to other BHs in the LHS, the formation and/or 
emission from the jet may be prevented for some unknown reasons.} \\
\indent Recently several sources have been observed to remain in the LHS during the whole
 duration 
of their outburst. These sources can either \correc{host} a NS (e.g. Aql X-1), or a BH(C) 
(XTE J1550$-$564, \SWIFT, GRO J0422+32, GRO J1719-24, and XTE J1118+480; 
\citet{sturner05,cadolle06a, brocksopp04}). Interestingly Aql X-1 and XTE J1550$-$564 
are also known to undergo major outbursts. The existence of LHS outbursts in XRBs brings 
interesting 
questions regarding the physics of accretion. In XTE J1550$-$564, \citet{sturner05} have 
suggested
that this outburst could correspond to a discrete accretion event. \citet{rodriguez06b} 
have suggested 
the same could also occur in Aql X-1. In both these cases, the outbursts had a short duration 
($\sim$30-40 days) compared to their previous outbursts. The similar duration of the 
outburst of
\IGR\ could by comparison indicate that a similar event took place. As pointed out 
by \citet{sturner05} 
the LHS outburst in GRO J0422+32, GRO J1719-24, and XTE J1118+480 could (by opposition to at 
least XTE J1550$-$564) be explained by a lower reservoir of material, these sources having 
a short orbital periods. \correc{Although we cannot know
whether IGR J17497-2821 will have a major outburst in the future, the fact that 
the companion of \IGR\ is a red giant K-type star \citep{paizis06} 
rules out a system with such a short orbital period.}

\begin{acknowledgements}
We are especially grateful to J. H. Swank and the RXTE mission planers 
for having 
accepted our ToO, and for their very rapid reaction to plan our observations. 
The Swift/BAT transient monitor results are kindly provided by the Swift/BAT team.
J.R. would like to thank S. Chaty for useful discussions.
This research has made use of data obtained through the High Energy 
Astrophysics Science Archive Center Online Service, provided by the 
NASA/Goddard Space Flight Center.
\end{acknowledgements}

%\bibliography{ms}

\begin{table}[htbp]
\caption{Journal of the \rxte\ observations analyzed in this paper. MJD 54000 is Sept. 22., 2006 
 The last two columns indicate the net count 
rates from PCA/PCU2 (3-30 keV) and HEXTE/ClusterB (18-200 keV).}
\begin{tabular}{cccccc}
\hline
Obs.& Obs. Id & MJD  & GTI & \multicolumn{2}{c}{Count rates}\\
\#  & (P92016) & (d)  & (s) & PCA & HEXTE \\
\hline
\hline
1 & 01-01-00 & 53998.2 & 6384 &  61.4 & 13.3\\
2 & 01-02-00 & 54000.2 & 3200 & 67.5 & 14.5\\
3 & 01-01-02 & 54001.3 & 3200 & 68.4& 14.3\\
4 & 01-02-01 & 54002.0 & 2368 & 66.7& 14.3\\
5 & 01-02-02 & 54003.2 & 3216  & 64.5& 15.3\\
6 & 01-01-01 & 54005.0 & 3232 & 60.8 & 13.0\\
7 & 01-03-00 & 54007.1 & 6480 &53.8& 12.2\\
\end{tabular}
\label{tab:log}
\end{table}
\begin{table}[htbp]
\caption{Best fit parameters obtained from the combined PCA+HEXTE spectra.
The spectral model consists of {\tt{phabs*edge*comptt}}. In all cases the reduced 
$\chi^2$ is around 1.
The seed photon temperature for Comptonization was frozen to 0.1 keV in all cases. 
Errors are given at the 90\% confidence level. $^\dagger$ $\times 10^{22}$ cm$^{-2}$.
$^\star$3-20 keV unabsorbed  $\times10^{-9}$\ergcms. }
\begin{tabular}{cccccc}
\hline
\hline
Obs.& N$_{\mathrm{H}}^\dagger$ & Edge& kT$_e$ & $\tau$ & flux$^\star$\\
   \#       &  & keV& keV & &  \\
\hline
1       & 4.1$\pm0.4$ & 6.8$\pm0.3$ & 37$_{-7}^{+118}$ & 1.6$_{-1.0}^{+0.3}$ & 1.0\\
2        & 3.8$\pm0.4$ & 6.6$\pm0.3$ & 34$_{-6}^{+21}$ & 1.8$_{-0.6}^{+0.3}$ & 1.14 \\
3        & 3.9$\pm0.4$ & 6.7$\pm0.3$ & 33$_{-5}^{+12}$ & 1.8$_{-0.4}^{+0.2}$ & 1.16\\
4        & 3.8$\pm0.5$ & 7.1$_{-0.3}^{+0.4}$ & 33$_{-5}^{+16}$ & 1.9$_{-0.5}^{+0.3}$  & 1.13\\
5       &3.3$\pm0.4$ & 6.8$_{-0.3}^{+0.4}$ & 37$_{-6}^{+25}$ & 1.7$_{-0.5}^{+0.3}$ & 1.08 \\
6       & 3.5$\pm0.4$ & 6.8$_{-0.3}^{+0.4}$ & 40$_{-8}^{+22}$ & 1.7$_{-0.5}^{+0.3}$ & 1.02 \\
7        &3.6$\pm0.4$ & 6.8$\pm0.3$ & 38$_{-7}^{+88.0}$ & 1.7$_{-1.1}^{+0.3}$ & 0.91 \\
\end{tabular}
\label{tab:fits}
\end{table}

\clearpage

\end{document}